\documentclass[11pt]{article}
\begin{document}
 \baselineskip 0.22in

\newtheorem{Thm}{Theorem}[section]
\newtheorem{Cor}[Thm]{Corollary}
\newtheorem{Lem}[Thm]{Lemma}
\newtheorem{Propn}[Thm]{Proposition}
\newtheorem{Def}[Thm]{Definition}
\newtheorem{rem}[Thm]{Remark}
\newtheorem{Asmp}[Thm]{Assumption}

\title{\textbf{Uniqueness Conditions for A Class of $\ell_0$-Minimization Problems}
\author{ Chunlei Xu\thanks{School of Mathematics, University of Birmingham, Edgbaston,
B15 2TT (email: {\tt cxx028@bham.ac.uk}). } ~ and Yun-Bin
Zhao\thanks{School of Mathematics, University of Birmingham,
Edgbaston, B15 2TT (email: {\tt y.zhao.2@bham.ac.uk}).  This author
was supported by the Engineering and Physical Sciences Research
Council (EPSRC) under the grant \#EP/K00946X/1, and partially
supported by the National Natural Science Foundation of China (NSFC)
under the grant \#11301016. }}}

\date{October 29, 2013}

 \maketitle

\textbf{Abstract.}  We consider a class of $\ell_0$-minimization
problems, which is to search for the partial sparsest solution to an
underdetermined linear system with additional constraints.   We
introduce several concepts, including $l_p$-induced norm ($0<p<1$),
maximal scaled spark and scaled mutual coherence, to develop several
new uniqueness conditions for the partial sparsest solution to this
class of $\ell_0$-minimization problems.   A further improvement of
some of these uniqueness criteria have been also achieved through
the so-called
concepts such as maximal scaled (sub)coherence rank.\\

\textbf{Key words}. $\ell_0$-minimization, uniqueness condition,
$l_p$-induced norm, scaled spark, scaled mutual (sub)coherence,
maximal scaled coherence rank.

\section{Introduction}
  Sparse representation, using only a few elementary atoms from a dictionary to represent data
(signals, images, etc.), has been widely used in engineering and
applied sciences recently (see, e.g., \cite{SM1998, TM2001,
LSLD2005, FNW2007, BED2009, ME2010,  DOLI2011, EK2012, ZL2012,
YZ2013} and the references therein). For a vector $x,$ let $\|x\|_0$
denote the `$\ell_0$-norm' of $x$, namely, the number of nonzero
components of $x.$
 In this paper, we consider the following model for the sparse
representation of the vector $b\in R^m: $
\begin{eqnarray}\label{PPPP}
\min \left\{\|x\|_0: ~M\left(\begin{array}{c} x\\
y \end{array}\right)=b,~y\in C \right\},
\end{eqnarray}
where $M=[A_1, A_2]\in R^{m\times (n_1+n_2)}, $ $m \leq n_1, $ is a
concatenation of $A_1\in R^{m\times n_1}$ and $A_2\in R^{m\times
n_2},$ and $C$ is a convex set in $R^{n_2}$ which can be interpreted
as certain constraints on the variable $y\in R^{n_2}.$ The solution
to the system
\begin{equation} \label{system}  M \left(\begin{array}{c} x\\
y
\end{array}\right)=b,  ~ y\in C, \end{equation}
  includes two parts: $x \in R^{n_1}$ and $y\in R^{n_2}. $ The
$\ell_0$-minimization problem (\ref{PPPP}) is to seek a solution
$z=(x,y)$ to the system (\ref{system}) such that the $x$-part is the
sparsest one, but there is no requirement on the sparsity of the
$y$-part of the solution. Such a sparsest solution $x$ can be called
the \emph{sparsest $x$-part solution} to the system  (\ref{system}).
The $\ell_0$-minimization problem is NP-hard (see Natarajan
\cite{N1995}), and can be called a partial $\ell_0$-minimization
problem, or partial sparsity-seeking problem.

Problem (\ref{PPPP}) is closely related to partial sparsity recovery
theory (see, e.g., Bandeira et al.\cite{BSV2013}, and Jacques
\cite{J2010}), and partial imaging reconstruction (Vaswani and Lu
 \cite{VL2010}), and the sparse Hessian recovery (Bandeira et al. \cite{BSV2012}).
 Problem (\ref{PPPP}) is general enough to include
several importance sparsity seeking problems as special cases. For
instance, the  $\ell_0$-minimization with inequality constraints:
$$ \min\{\|x\|_0:  ~ Ax\geq b\} $$
can be written as
$$ \min\{\|x\|_0: ~ Ax-y \geq b, ~y\geq 0\}, $$
which is a special case of (\ref{PPPP}). Moreover, the normal
$\ell_0$-minimization
\begin{equation}  \label{OOOO} \min \{\|x\|_0:
Ax=b\}
\end{equation}
is also an important special case of (\ref{PPPP}). In fact, Problem
(\ref{PPPP}) is reduced to (\ref{OOOO}) when $A_2=0. $

 The uniqueness of
the standard $\ell_0$-minimization (\ref{OOOO}) has been widely
investigated, and has been established by using the so-called spark
of a matrix $A$ (see Donoho and Elad \cite{DE2003}),  denoted by
$\textrm{Spark} (A), $  which is the smallest number of columns of a
matrix that are linearly dependent.  It was shown in \cite{DE2003,
DDEK2012, BED2009}] that for a given linear system $Ax=b$, if there
exists a solution $x$ satisfying $ \|x\|_0<
\frac{1}{2}\textrm{Spark}(A), $ then $x$ is necessarily the unique
sparsest solution to (\ref{OOOO}).

Since the computation of spark is generally intractable, some other
verifiable conditions have been developed in the literature. For
instance, by the mutual coherence \cite{DH2001}, the largest
absolute value of inner products between different normalized
columns of $A$, i.e.,
$$
\mu(A)=\max_{1\leq i\not= j \leq n}  \frac{|\langle a_i,
a_j\rangle|} {\|a_i\|_2\cdot\|a_j\|_2},
$$
where $a_i$ is the $i$-th column in $A$, $i=1,\cdots,n$.  The mutual
coherence gives a computable lower bound for the spark
\cite{DE2003}, i.e., $ \textrm{Spark}(A)\geq 1+\frac{1}{\mu(A)}, $
which yields following uniqueness condition (see, e.g.,
\cite{DE2003, DDEK2012, BED2009}):  For a given linear system
$Ax=b$, if there exists a solution $x$ satisfying $ \|x\|_0<
\frac{1}{2}(1+\frac{1}{\mu(A)}), $ then $x$ is necessarily the
unique sparsest solution to (\ref{OOOO}). However, the mutual
coherence condition might be very restrictive in some situations,
and fails to provide a good lower bound for the spark.  For
instance, when $A=[I_{m\times m},e]\in R^{m\times {m+1}}$ with
$e=[1,1,\cdots,1]^T\in R^m$, we have
$$\textrm{Spark}(A)=m+1, ~~ 1+\frac{1}{\mu(A)}=2. $$ In order to
improve the lower bound for spark, Zhao \cite{YZ2013} has introduced
the concept of coherence rank, submutual coherence and scaled mutual
coherence, and has developed several new and improved uniqueness
sufficient conditions for the  solution to $\ell_0$-minimization
(\ref{OOOO}).

 So far, the uniqueness of the sparsest
$x$-part solution  to the general sparsity module (\ref{PPPP})
  has not well developed.  The main purpose of this paper
is to study such uniqueness  and to establish some criteria under
which the problem (\ref{PPPP}) has a unique sparsest $x$-part
solution. These results will be established through  some new
concepts such as the $l_p$-induced norm, the (maximal) scaled spark,
coherence, and coherence rank associated with a pair of matrices
$(A_1\in R^{m\times n_1}, A_2\in R^{m\times n_2}).$
 These concepts can be seen as a generalization of those in
 \cite{YZ2013}.

This paper is organized as follows. In Section 2, we will develop
sufficient conditions for the uniqueness of $x$-part solutions to
the $\ell_0$-minimization problem (\ref{PPPP}) in terms of $l_p$-
induced norm, and  such concepts as maximal scaled spark, and
minimal or maximal scaled mutual coherence. A further improvement of
these conditions is provided in Section 3.

\section{Uniqueness criteria for the $\ell_0$-minimization problem (\ref{PPPP})}

The uniqueness of the sparsest $x$-part solution to the system
(\ref{system})  can be developed through different concepts and
properties of matrices. One of such important concept is
\textit{spark}  together with its variants,
 which provides a connection between the null space of a matrix
 and the sparsest solution to linear equations.  In this section, we show that  the method used for
 developing uniqueness claims for the $\ell_0$-minimization (\ref{OOOO}) can be used for the development of similar claims to
  the system (\ref{system}), while the extra variable $y$ in the system
$M\left(\begin{array}{c} x\\ y \end{array}\right)=b$
  increases the complexity of the (\ref{PPPP}) problem.  Our first sufficient uniqueness condition for
  the sparsest $x$-part
  solution to (\ref{PPPP}) can be developed
 by using the so-called $l_p$-induced norm, as shown in the following subsection.

\subsection{A uniqueness condition based the $l_p$-induced norm}
For any $0<p<\infty$ and a vector $x\in R^n,$  let $\|x\|_p =
\left(\sum_{i=1}^n|x_i|^p\right)^{1/p}.$  When $p\in (0,1),$
$\|x\|_p$ is called the $l_p$ quasi-norm of $x.$ We now introduce
the following concept.

\begin{Def}   For any given
matrix $A\in R^{m\times n}$, when $0<p<1$, the $l_p$-induced norm of
  $A,$ denoted by $\psi_p(A),$  is defined by
\begin{equation}\label{eqn1.7}
\psi_p(A)= \sup_{0\not=z\in
R^n}\frac{\|Az\|_p^p}{\|z\|_p^p}=\sup_{\|z\|_p^p\leq1}\|Az\|_p^p.
\end{equation}

\end{Def}

Clearly, for a fixed $p \in (0,1),$  $\psi_p(A) $ is a norm since it
satisfies the following properties: $ \psi_p(A) \geq 0,$  $
\psi_p(A) > 0$ for any $A\not =0,$    $ \psi_p(\alpha A) = \alpha
\psi_p(A)$ for any $\alpha \geq 0,$ and $ \psi_p(A+B) \leq \psi_p(A)
+\psi_p(B)$ for any matrices $A, B $ with same dimensions. It is
worth mentioning that the triangle inequality above follows from the
 property: $\|x+y\|_p^p  \leq \|x\|_p^p +\|x\|_p^p $ (see,
e.g., \cite{FPRU2010}).    Note that, for every entry $z_i$, as $p$
tends to zero, $|z_i|^p$ approaches to 1 for $z_i\neq 0$ and 0 for
$z_i=0$. Thus for any given $ z\in R^n, $ we have
\begin{equation} \label{555}
 \lim_{p\rightarrow 0}\|z\|_p^p=\lim_{p\rightarrow
0}\sum_{i=1}^n|z_i|^p= \|z\|_0,
\end{equation}
which indicates that the  `$\ell_0$-norm' $\|z\|_0$ can be
approximated by $\|x\|_p^p$ with sufficiently small $p\in (0,1). $
Note that for a given matrix $A$, $\psi_p(A) $ is continuous with
respect to $p \in (0,1). $ Thus there might exists a positive number
$\eta$ such that $\eta= \lim_{p\to 0^+} \psi_p(A).$   We  assume
that the following property holds for the matrix $M=[A_1, A_2]$ when
$p$ tends to 0.

\begin{Asmp} \label{Asmp1.1} Assume that matrices $A_1, A_2$
satisfy the following properties: (i) $A_2^T A_2$ is a nonsingular
matrix, and (ii) there exists a positive constant, denoted by
$\psi_0(A_2^\dag A_1)),$ such that
$$  \psi_0(A_2^\dag A_1))  = \lim_{p\to 0^+} \psi_p(A_2^\dag A_1)),$$
where $A_2^\dag =(A^T_2 A_2)^{-1} A_2^T ,$  the pseudo-inverse of
$A_2.$
\end{Asmp}

Under the Assumption \ref{Asmp1.1} and  by (\ref{eqn1.7}) and
(\ref{555}), we immediately have the following  inequality:
\begin{equation}\label{eqn1.18}
\| (A_2^\dag A_1) z\|_0 =\lim_{p\to 0_+} \| (A_2^\dag A_1) z\|^p_p
\leq \lim_{p\to 0_+} (\psi_p (A)\|z\|^p_p ) \leq \psi_0(A_2^\dag
A_1) \|z\|_0
\end{equation} for any
$ z\in R^n.$  We now state a uniqueness condition for Problem
(\ref{PPPP}) under Assumption \ref{Asmp1.1}.

\begin{Thm} \label{Thmlpinducenorm}
Consider the system (\ref{system})  with $A_1\in R^{m\times n_1},$
$A_2\in R^{m\times n_2},$ and $ m<n_1. $ Let Assumption
\ref{Asmp1.1} be satisfied. Then if there exists a solution $(x,y)$
to  the system (\ref{system}) satisfying that
\begin{eqnarray}\label{eqn1.8}
\|x\|_0<\frac{1}{2}\frac{Spark(M)}{(1+\psi_0(A_2^\dag A_1))},
\end{eqnarray}
 $x$ must be  the unique
 sparsest x-part solution to the system (\ref{system}).
\end{Thm}

{\it Proof.} Assume the contrary that there is another solution
$(x^{(1)},y^{(1)})$ to the system (\ref{system}) such that $x^{(1)}$
is the sparsest
$x$-part  and $  x^{(1)}  \neq  x $ and $\|x^{(1)}\|\leq \|x\|_0<\frac{1}{2}\frac{\textrm{Spark}(M)}{(1+\psi_0(A_2^\dag A_1))}$. Since both $(x,y)$ and $(x^{(1)}, y^{(1)})$ are  solutions to the linear system $M\left(\begin{array}{c} x\\
y
\end{array}\right)=b$, we have
\begin{eqnarray}\label{eqn1.9}
A_1(x-x^{(1)})+A_2(y-y^{(1)})=0.
\end{eqnarray}
Since $A_2^TA_2$ is nonsingular, $y-y^{(1)}$ can be uniquely
determined by $x-x^{(1)}$, i.e.,
\begin{eqnarray}\label{eqn1.10}
y^{(1)}-y=A_2^\dag A_1(x-x^{(1)}),
\end{eqnarray}
where $A_2^\dag$ is the pseudo-inverse of $A_2$ given by $A_2^\dag=
(A_2^TA_2)^{-1}A_2^T.$  From
(\ref{eqn1.9}), we know that $\left(\begin{array}{c} x-x^{(1)} \\
y-y^{(1)}\end{array} \right)$ is in the null space of the matrix $M
=[A_1, A_2]$. This implies that the $\textrm{Spark}(M)$ is a lower
bound for $\left\|\left(\begin{array}{c} x-x^{(1)} \\
y-y^{(1)}\end{array} \right)\right\|_0$, i.e.,
\begin{eqnarray}\label{eqn1.11}
\|x-x^{(1)}\|_0+\|y-y^{(1)}\|_0=\left\|\left(\begin{array}{c} x-x^{(1)} \\
y-y^{(1)}\end{array} \right) \right\|_0\geq \textrm{Spark}(M).
\end{eqnarray}
Substituting (\ref{eqn1.10}) into (\ref{eqn1.11}) leads to
\begin{eqnarray}\label{eqn1.12}
\|x-x^{(1)}\|_0+\|A_2^\dag A_1(x-x^{(1)})\|_0\geq \textrm{Spark}(M).
\end{eqnarray}
Under Assumption \ref{Asmp1.1},  one has
$$\|A_2^\dag A_1(x^{(1)}-x_2)\|_0\leq\psi_0(A_2^\dag
A_1)\cdot\|x-x^{(1)}\|_0.$$ Merging (\ref{eqn1.12}) and the
inequality above leads to
$$
(1+\psi_0(A_2^\dag A_1))\|x-x^{(1)}\|_0 \geq   \textrm{Spark}(M),$$
Therefore,
$$  2\|x\|_0 \geq \|x^{(1)}\|_0+\|x\|_0\geq\|x-x^{(1)}\|_0  \geq
\frac{\textrm{Spark}(M)}{1+\psi_0(A_2^\dag A_1)}.
$$
Thus $\|x\|_0\geq
\frac{1}{2}\frac{\textrm{Spark}(M)}{(1+\psi_{A_2^\dag A_1}(0))},$
contradicting with (\ref{eqn1.8}). Therefore $x $ must be the unique
 sparsest $x$-part solution to  Problem (\ref{PPPP}).

The above result provides a new uniqueness criteria for the problem
(\ref{PPPP}) by using $l_p$-induced norm.  However, the above
analysis relies on the nonsingularity of $A_2^TA_2$ which might not
be satisfied in more general situations.  Thus we need to develop
some other uniqueness criteria for the problem from other
perspectives.

\subsection{Uniqueness based on scaled spark and scaled mutual coherence}

In this section, we develop uniqueness conditions for Problem
(\ref{PPPP}) by using the so-called scaled spark and scaled mutual
coherence.

\begin{Lem} [\cite{BED2009}]\label{Lem1.1}  For any matrix $M$ and any scaling matrix $W$, one has
$$
Spark(WM)\geq1+\frac{1}{\mu (WM)}.
$$
\end{Lem}

In the remainder of this paper, we use ${\cal N} (\cdot) $ to denote
the null space of a matrix. Our first uniqueness criterion based on
scaled spark is given as follows.

\begin{Thm}   \label{UniScaledSpark}
Consider the system  (\ref{system})   where  $A_1\in R^{m\times
n_1},$  $A_2\in R^{m\times n_2}$ and $ m<n_1. $ If there exists a
solution $(x, y)$ to the system (\ref{system}) satisfying
\begin{eqnarray}\label{eqn1.14}
\|x\|_0<\frac{1}{2}Spark(B^TA_1),
\end{eqnarray}
where $B$ is a basis of $\mathcal{N}(A_2^T)$, then $x$ is the unique
sparsest $x$-part solution to the  system (\ref{system}).

\end{Thm}

{\it Proof.} Assume that $(x^{(1)},y^{(1)}) \not =(x,y) $ is a
solution   to the system (\ref{system}) satisfying that $x^{(1)}\neq
x $, and $\|x^{(1)}\|_0 \leq \|x
\|_0<\frac{1}{2}\textrm{Spark}(B^TA_1), $ where $B$ is a basis of
${\cal N} (A_2^T).$  Note that $\left(\begin{array}{c}x^{(1)}-x\\
y^{(1)}-y
\end{array}\right) $ is in the null space of $M= [A_1,A_2]$, so
\begin{equation}\label{eqn1.15}
A_1(x^{(1)}-x_2)=-A_2(y^{(1)}-y).
\end{equation}
Note that the range space of $A_2$ is orthogonal to the null space
of $A_2^T$, namely,
 $$\mathcal{R}(A_2)=\mathcal{N}(A_2^T)^\bot.$$
Let $B$ be an arbitrary  basis of $\mathcal{N}(A_2^T).  $ Since the
right-hand side of  (\ref{eqn1.15}) is in  $\mathcal{R}(A_2), $ by
multiplying both sides of the equation (\ref{eqn1.15}) by $B^T$, we
get
$$B^TA_1(x^{(1)}-x)=0, $$ which implies that
\begin{equation}\label{eqn1.19}
\|x^{(1)}-x\|_0\geq \textrm{Spark}(B^TA_1). \end{equation}
Therefore,
$$ 2\|x|_0 \geq    \|x^{(1)}\|_0+\|x\|_0\geq \textrm{Spark}(B^TA_1).
$$
i.e., $\|x\|_0\geq \frac{1}{2}\textrm{Spark}(B^TA_1), $ leading to a
contradiction.  Therefore, the system (\ref{system}) has a unique
sparsest $x$-part solution.  \\

Let $F$ be a set of all bases of  ${\cal N} (A_2^T),  $ namely,
\begin{eqnarray*}
F=\{B\in R^{m\times q}:  ~B \textrm{ is a basis of
}~\mathcal{N}(A_2^T)\},
\end{eqnarray*}
where $q$ is the dimension of $\mathcal{N}(A_2^T)$.\\

 From the
definition of the spark, we know that $\textrm{Spark}(B^TA_1)$ is
bounded. Hence, there exists the supremum of
$\textrm{Spark}(B^TA_1)$, defined as follows.

\begin{Def}
For any matrix $A_1\in R^{m\times n_1}$ with $m<n_1$, let
\begin{eqnarray}\label{eqn1.20}
\textrm{Spark}_{A_2}^*(A_1)=\sup_{B \in F} \textrm{Spark}(B^TA_1)  .
\end{eqnarray}
$\textrm{Spark}_{A_2}^*(A_1)$ is called the  maximal scaled spark of
$A_1 $ over $F$ (the set of bases of ${\cal N}(A_2)).$
\end{Def}

The inequality (\ref{eqn1.19}) in the proof of Theorem
\ref{UniScaledSpark} holds for all bases $B$ of
$\mathcal{N}(A_2^T)$.  Therefore, the spark condition
(\ref{eqn1.14}) can be further enhanced by using
$\textrm{Spark}_{A_2}^*(A_1). $

\begin{Thm} \label{UniONSSpark}
Consider the system (\ref{system}) where $A_1\in R^{m\times n_1}$
and $A_2\in R^{m\times n_2}$  and  $ m<n_1. $ If there exists a
solution $(x,y)$ to (\ref{system}) satisfying
\begin{eqnarray}\label{eqn1.21}
\|x\|_0<\frac{1}{2}\textrm{Spark}_{A_2}^*(A_1),
\end{eqnarray}
where $\textrm{Spark}_{A_2}^*(A_1)$ is  given by  (\ref{eqn1.20}),
then $x$ is the unique sparsest $x$-part solution to (\ref{system}).

\end{Thm}

From Lemma \ref{Lem1.1},   the scaled mutual coherence may provide a
lower bound for the scaled spark. An immediate consequence of
  Theorem \ref{UniScaledSpark} is the   corollary below.

\begin{Cor}   \label{CorUniScaledMC}
For a given system (\ref{system}) where $A_1\in R^{m\times n_1}, $
$A_2\in R^{m\times n_2}$ and  $ m<n_1. $ If there exists a solution
$(x,y)$ to (\ref{system}) satisfying
\begin{eqnarray}\label{eqn1.22}
\|x\|_0<\frac{1}{2}\left(1+\frac{1}{\mu(B^TA_1)}\right),
\end{eqnarray}
where $B$ is a basis of $\mathcal{N}(A_2^T),$ then $x$ is the unique
 sparsest $x$-part solution to the system (\ref{system}).
\end{Cor}

Note that Corollary \ref{CorUniScaledMC} holds for any bases $B$ of
$\mathcal{N}(A_2^T)$.   So it makes sense to further enhance the
bound (\ref{eqn1.22}) by introducing the following definition.

\begin{Def} For any matrix $A_1\in R^{m\times n_1}$ ($m<n_1$) and $A_2\in R^{m\times n_2},$ let
\begin{eqnarray}\label{eqn1.23}
~\mu_{A_2}^*(A_1) =\min_{B\in F} \mu(B^TA_1), ~~
\mu^{**}_{A_2}(A_1) =\max_{B\in F} \mu(B^TA_1).
\end{eqnarray}
  $\mu_{A_2}^*(A_1)$ is called the minimal scaled coherence of $A_1$ over
  $F,$ and  $\mu^{**}_{A_2}(A_1)$  is called the  maximal scaled coherence of $A_1$ over
  $F. $
\end{Def}

Based on Lemma \ref{Lem1.1} and the above definition,  we have  the
following result.

\begin{Lem}\label{Lem1.2}  For any basis $B$ of $\mathcal{N}(A_2^T),$
we have
\begin{eqnarray}\label{eqn1.24}
1+\frac{1}{\mu(B^TA_1)}\leq1+\frac{1}{\mu_{A_2}^*(A_1)}\leq
\textrm{Spark}_{A_2}^*(A_1),
\end{eqnarray}

\end{Lem}

{\it Proof.} The first inequality holds by the definition of
$\mu_{A_2}^*(A_1) .$  From Lemma \ref{Lem1.1}, we have
\begin{eqnarray*}
1+\frac{1}{\mu(B^TA_1)}\leq \textrm{Spark}(B^TA_1)~~\textrm{ for
every basis } B \textrm{ of } \mathcal{N}(A_2^T).
\end{eqnarray*}
By  (\ref{eqn1.21}),  we have that $ \textrm{Spark}(B^TA_1) \leq
\textrm{Spark}_{A_2}^*(A_1), $ thus
\begin{eqnarray*}
1+\frac{1}{\mu(B^TA_1)} \leq \textrm{Spark}_{A_2}^*(A_1) ~~\textrm{
for all } B\in F.
\end{eqnarray*}
Since the right-hand side of the above  is  fixed, which is an upper
bound for the left-hand side for any $B\in F,$ we conclude that
$$
\textrm{Spark}_{A_2}^*(A_1)  \geq   \max_{B\in F}
\left\{1+\frac{1}{\mu(B^TA_1)}\right\} = 1+\frac{1}{\min_{B\in
F}\{\mu(B^TA_1)\}}= 1+\frac{1}{\mu_{A_2}^*(A_1)}.
$$

By Theorem \ref{UniONSSpark} and Lemma \ref{Lem1.2}, we have the
next enhanced uniqueness claim.

\begin{Thm}  \label{Thm1.4} For a given system (\ref{system})
with $A_1\in R^{m\times n_1},$ $A_2\in R^{m\times n_2}$ and $ m<n_1.
$ If there exists a solution $(x,y)^T$ satisfying
\begin{eqnarray}\label{eqn1.25}
\|x\|_0<\frac{1}{2}\left(1+\frac{1}{\mu_{A_2}^*(A_1)}\right),
\end{eqnarray}
where $\mu_{A_2}^*(A_1)$ is the minimal scaled coherence  of $A_1$
on $F, $ then $x $ is the unique  sparsest $x$-part solution to the
system (\ref{system}).
\end{Thm}

 \textbf{Remark.} The uniqueness criteria established in this section can be seen as
 certain generalization of that of  sparsest solutions to systems of linear equations.
 For instance,   when $ A_2= 0, $
   the null space of $A_2^T$ is the whole space $R^m$. Hence, by letting
$B=I$, the corresponding scaled mutual coherence
and scaled spark become
$$
 \mu(B^TA_1)=\mu(A_1), ~~ \textrm{spark}(B^TA_1)=\textrm{spark}(A_1),
$$   The results in this section are reduced to existing ones \cite{DE2003, DDEK2012, BED2009}.
It is worth noting that the spark type uniqueness conditions
 are derived from the property of null spaces.  It is worth mentioning that  the null space based analysis is not the unique way to derive
 uniqueness criteria for sparsest solutions. Some other approaches such as the so-called range space
 property  (see, e.g., \cite{YZ2013,ZL2012}) and orthogonal
  projection from $R^{n_1+n_2}$ to $\mathcal{N}(A_2^T)$ \cite{BSV2013}  can be also used to develop uniqueness
  criteria.

\section{Further Improvement of some uniqueness conditions}

Since  spark conditions are difficult to verify, the mutual
coherence conditions play an important role in the uniqueness theory
for  the $\ell_0$-minimization problem (\ref{system}). As shown in
Lemma \ref{Lem1.2}, $1+\frac{1}{\mu_{A_2}^*(A_1)}$ is a good lower
bound for $\textrm{spark}^*_{A_2}(A_1) $ which is an improved
version of the bound (\ref{eqn1.22}).  In this section, we aim to
  enhance the uniqueness claim (\ref{eqn1.25}) by further improving
the lower bound of $\textrm{Spark}^*_{A_2}(A_1)$ under some
situations. Following the discussions in \cite{YZ2013}, we introduce
the so-called scaled coherence rank, scaled subcoherence and scaled
sub-coherence rank to achieve certain improvement on uniqueness
conditions developed in section 2.

\subsection{Maximal (sub) coherence and rank}

 Let us first recall several concepts which were introduced by Zhao
\cite{YZ2013}.  For a given matrix $A\in R^{m\times n}$ with columns $a_i,  i=1, ..., n,$ consider
the index set
$$ S_i(A): = \left\{j:  ~ j\not =i, ~ \frac{|a_i^T a_j|}{\|a_i\|_2\cdot \|a_j\|_2}
=\mu(A)  \right\}, ~i=1, ..., m. $$    Let  $\alpha_i(A)$ be the
cardinality of $S_i(A), $ and $\alpha (A) $ be the largest one among
$\alpha_i(A)$'s, i.e., $$  \alpha (A) = \max_{1\leq i\leq m}
\alpha_i (A) = \max_{1\leq i\leq m} |S_i (A)| . $$  $\alpha (A)$  is
called the coherence rank of
 $A. $ Let $i_0$ be an
index such that $ \alpha(A)= \alpha_{i_0} (A) =|S_{i_0}(A)|. $
Define  $$ \beta  (A) =\max_{1\leq i\leq m, ~i\not= i_0} \alpha_i(A)
= \max_{1\leq i\leq m, ~i\not= i_0} |S_i(A)|, $$  which is called
the sub-coherence rank of $A.$  Also we define by
$$ \mu^{(2)} (A)= \max_{i\not= j}
\left\{\frac{|a_i^T a_j|}{\|a_i\|_2 \cdot \|a_j\|_2}:
  ~ \frac{|a_i^T a_j|}{\|a_i\|_2 \cdot \|a_j\|_2} < \mu (A)  \right\},  $$
 the second largest absolute value of the inner product
between two normalized columns of $A.$ $\mu^{(2)}(A)$ is called the
sub-mutual coherence of $A. $\\

Consider the submutual coherence $\mu^{(2)}(B^TA_1)$ with a scaling
matrix $B\in F.$ We introduce the following new concept.

\begin{Def} Let   $A_1\in R^{m\times n_1}$ ($m<n_1$) and $A_2\in R^{m\times n_2}$ be two matrices,
 and $F$ is the set of bases of $
{\cal N} (A^T_2). $

(i) The maximal scaled submutual coherence  of $A_1$ on $F,$ denoted
by
 $\mu^{**(2)}_{A_2}(A_1)$, is defined as
\begin{equation}\label{2211}
\mu^{**(2)}_{A_2}(A_1)=\sup_{B\in F} \mu^{(2)}(B^TA_1) .
\end{equation}

(ii)  The maximal scaled coherence rank of $A_1$ on $F,$ denoted by
$\alpha_{A_2}^*(A_1)$, is defined as
\begin{equation}\label{2222}
\alpha_{A_2}^*(A_1)=\sup_{B\in F}\{\alpha(B^TA_1)\}.
\end{equation}

(iii)  The maximal scaled subcoherence rank on $F,$ denoted by
$\beta^*_{A_2}(A_1)$, is defined as
\begin{equation}\label{2233}
\beta_{A_2}^*(A_1)=\sup_{B\in F}\{\beta(B^TA_1)\}.
\end{equation}
\end{Def}

It is easy to see the following relationship between
$\alpha(B^TA_1)$,
 $\beta(B^TA_1)$, $\alpha_{A_2}^*(A_1)$ and $\beta_{A_2}^*(A_1): $
For every basis $B$ of $\mathcal{N}(A_2^T),$ we have
\begin{equation} \label{alpha-beta}
1\leq\beta(B^TA_1)\leq\alpha(B^TA_1)\leq\alpha_{A_2}^*(A_1)
~\textrm{and}~1\leq\beta(B^TA_1)\leq\beta_{A_2}^*(A_1)\leq\alpha_{A_2}^*(A_1).
\end{equation}


\subsection{Improved lower bounds of $spark_{A_2}^*(A_1)$}

Following the  method used to improve the lower bound of $Spark(A)$
in \cite{YZ2013}, we can find an enhanced lower bound of
$\textrm{Spark}_{A_2}^*(A_1)$ via the concepts introduced in Section
3.1. We will make use of the following two lemmas.

\begin{Lem} (Brauer \cite{AB1946}) \label{Thm1.5}
For any matrix $A\in R^{n\times n}$ with $n\geq 2$, if $\lambda$ is an eigenvalue of $A$, there is a pair $(i,j)$ of positive integers with $i\neq j$ ($1\leq i,j\leq n$) such that
\begin{eqnarray*}
|\lambda-a_{ii}|\cdot|\lambda-a_{jj}|\leq\Delta_i\Delta_j,
\end{eqnarray*}
where $\Delta_i:=\sum^n_{j=1,j\neq i}|a_{ij}|$ for $1\leq i\leq n.$
\end{Lem}

Merging Theorem 2.5 and Proposition 2.6 in \cite{YZ2013} yields the
following result.

\begin{Lem}(Zhao \cite{YZ2013}) \label{Thm25inZhao}
Let $A\in R^{m\times n},$ and let $\alpha(A)$ and $\beta(A)$ be the
coherence rank and subcoherence rank of $A$, respectively.
Suppose that one of the following conditions holds: $(i)  ~~
\alpha(A)<\frac{1}{\mu(A)};  ~~~ (ii) ~
\alpha(A)\leq\frac{1}{\mu(A)}\textrm{ and } \beta(A)<\alpha(A). $
Then $\mu^{(2)}(A)>0$ and
\begin{eqnarray*}
Spark(A) & \geq &
1+\frac{2[1-\alpha(A)\beta(A)\bar{\mu}(A)^2]}{\mu^{(2)}(A)\{\bar{\mu}(A)(\alpha(A)+\beta(A))+\sqrt{\bar{\mu}(A)^2
(\alpha(A)-\beta(A))^2+4}\}}  \\
& > & 1+\frac{1}{\mu(A)}
\end{eqnarray*}
where $\bar{\mu}(A)=\mu(A)-\mu^{(2)}(A)$ and  $\mu^{(2)}(A)$ is the subcoherence of $A$.
\end{Lem}

Based on Lemma \ref{Thm25inZhao}, we can construct an enhanced lower
bound of $\textrm{Spark}_{A_2}^*(A_1)$ under some conditions, in
terms of the scaled coherence rank and   scaled subcoherence rank.

\begin{Thm}\label{Thmsparkstar1}
Consider the system (\ref{system}) where $A_1\in R^{m\times n_1},$
$A_2\in R^{m\times n_2}$ and $ m<n_1. $ Suppose that one of the
following conditions holds: (i)
  $\alpha(B^TA_1)<\frac{1}{\mu(B^TA_1)}\textrm{ for all } B\in F$;
 ~ (ii)  $\alpha(B^TA_1)\leq\frac{1}{\mu(B^TA_1)}$ and $\beta(B^TA_1)<\alpha(B^TA_1)\textrm{ for all } B \in F.$
Then for any $B\in F$, we have that $\mu^{(2)}(B^TA_1)>0$ and
 \begin{eqnarray*}
Spark_{A_2}^*(A_1) & \geq & \sup_{B\in F}
\bigg\{1+\frac{2[1-\alpha(B^TA_1)\beta(B^TA_1)\bar{\mu}(B^TA_1)^2]}{\mu^{(2)}(B^TA_1)\{\bar{\mu}(B^TA_1)
(\alpha(B^TA_1)+\beta(B^TA_1))+\sqrt{\Delta}\}}\bigg\}\\
& \geq  & 1+\frac{1}{\mu^*_{A_2}(A_1)}.
\end{eqnarray*}
where $\bar{\mu}(B^TA_1)=\mu(B^TA_1)-\mu^{(2)}(B^TA_1)$ and
 $\Delta=[\bar{\mu}(B^TA_1)]^2(\alpha(B^TA_1)-\beta(B^TA_1))^2+4$.

\end{Thm}

{\it Proof.} Under conditions (i) and (ii),  by Lemma
\ref{Thm25inZhao},  for any $B\in F$ we have that
$\mu^{(2)}(B^TA_1)>0$ and
\begin{eqnarray}\label{eqn1.33}
\textrm{Spark}(B^TA_1) & \geq  & \varphi(B^TA_1)  \nonumber \\ & =:
&
1+\frac{2[1-\alpha(B^TA_1)\beta(B^TA_1)\bar{\mu}(B^TA_1)^2]}{\mu^{(2)}(B^TA_1)\{\bar{\mu}(B^TA_1)(\alpha(B^TA_1)+\beta(B^TA_1))+\sqrt{\Delta}\}},
\end{eqnarray}
  where
$\bar{\mu}(B^TA_1)=\mu(B^TA_1)-\mu^{(2)}(B^TA_1)$ and
 $\Delta=[\bar{\mu}(B^TA_1)]^2(\alpha(B^TA_1)-\beta(B^TA_1))^2+4$. The above inequality holds for any basis $B\in
 F.$
By the definition of $\textrm{Spark}^*_{A_1}(A_1),$  we have
\begin{eqnarray*}
\textrm{Spark}^*_{A_2}(A_1)\geq \textrm{Spark}(B^TA_1)\textrm{ for
any  }B\in F.
\end{eqnarray*}
Thus it follows from  (\ref{eqn1.33}) that
\begin{small}\begin{eqnarray}\label{eqn1.34}
\textrm{Spark}^*_{A_2}(A_1)\geq  \varphi(B^TA_1) ~ \textrm{ for all
}B\in F
\end{eqnarray}\end{small}
Inequality (\ref{eqn1.34}) implies that the value of
$\varphi(B^TA_1)$ is bounded by the constant
$\textrm{Spark}^*_{A_2}(A_1)$. Hence, the superimum of
$\varphi(B^TA_1)$ over  $ F$ should be bounded by
$\textrm{Spark}^*_{A_2}(A_1)$, namely,
\begin{eqnarray*}
\textrm{Spark}^*_{A_2}(A_1)\geq\sup_{B\in F} \varphi(B^TA_1) .
\end{eqnarray*}
By Lemma \ref{Thm25inZhao} again, under conditions (i) and (ii), we
see that $\varphi(B^TA_1)>  1+\frac{1}{\mu(B^TA_1)}$. Therefore, the
superimum of $\varphi(B^TA_1)$ should be greater than the value of
$1+\frac{1}{\mu(B^TA_1)}$ for any basis $B \in F$, i.e.,
\begin{eqnarray*}
\sup_{B\in F} \varphi(B^TA_1) >1+\frac{1}{\mu(B^TA_1)} \textrm{ for
any $B \in F$}.
\end{eqnarray*}
This in turn implies  that
$$
\sup_{B\in F}\{\varphi(B^TA_1)\} \geq  \max_{B\in F}
\left\{1+\frac{1}{\mu(B^TA_1)} \right\} =
1+\frac{1}{\mu^*_{A_2}(A_1)},
$$
where the last equality follows from the definition of
$\mu^*_{A_2}(A_1)$.  Therefore, under conditions (i) and (ii), we
conclude that
$$
\textrm{spark}^*_{A_2}(A_1) \geq \sup_{B\in
F}\{\varphi(B^TA_1)\}\geq 1+\frac{1}{\mu^*_{A_2}(A_1)},
$$
as claimed.\\

Conditions (i) and (ii)  in Theorem  \ref{Thmsparkstar1} rely on
$B\in F. $ A similar conditions without relying on $B$ can be also
established as shown by the next result.

\begin{Thm}\label{Thmsparkstar}
Consider the system (\ref{system}) with $A_1\in R^{m\times n_1},$
$A_2\in R^{m\times n_2}$ and $ m<n_1. $ Let $ \mu^{**}_{A_2}(A_1), $
and $ \mu^{**(2)}_{A_2}(A_1), \alpha_{A_2}^*(A_1),
\beta^*_{A_2}(A_1)$ are four constants defined by (\ref{eqn1.23}),
(\ref{2211})-(\ref{2233}), respectively. Suppose that one of the
following conditions holds:
 (i) $\alpha_{A_2}^*(A_1)<\frac{1}{\mu^{**}_{A_2}(A_1)}$;
 (ii) $\alpha_{A_2}^*(A_1)\leq\frac{1}{\mu^{**}_{A_2}(A_1)} \textrm{
and}~\beta^*_{A_2}(A_1)<\alpha^*_{A_2}(A_1).$ Then
$\mu^{**(2)}_{A_2}(A_1)>0$ and
\begin{eqnarray*}\label{eqnsparkstar}
\textrm{Spark}^*_{A_2}(A_1)\geq \varphi^*=
1+\frac{\sqrt{\rho}-(\alpha^*_{A_2}(A_1)+\beta^*_{A_2}(A_1))\bar{\mu}^*}{2\mu_{A_2}^{**(2)}(A_1)},
\end{eqnarray*}
where $\bar{\mu}^*=\mu^{**}_{A_2}(A_1)-\mu_{A_2}^{**(2)}(A_1)$ and
$\rho=\left(\alpha_{A_2}^* (A_1)-\beta^*_{A_2} (A_1) \right)^2
(\bar{\mu}^*)^2+4.$
\end{Thm}

{\it Proof.} Note that $ \alpha (B^TA_1) \in \{0, 1, ..., n\}$ for
any $B\in F.$  By the definition of $\alpha^*_{A_2}(A_1)$ which is
the maximum value of $\alpha (B^TA_1)$ over $F,$   this maximum is
attainable, that is, there exists  a $\widehat{B}\in F$ such that
$$ \alpha^*_{A_2}(A_1) = \alpha (\widehat{B}^TA_1).$$
For such a basis $\widehat{B}\in F$, without loss of generality, we
assume that all columns of $\widehat{B}^TA_1$ are normalized in the
sense that the $l_2$-norm of every column  of $\widehat{B}^TA_1$ is
1. Note also that the spark, mutual coherence, subcoherence,
coherence rank, and subcoherence rank are invariant under
normalization.

Let $p=\textrm{Spark}(\widehat{B}^TA_1)$ and $\{c_1,\cdots,c_p\}$ be
the set of $p$ columns from $\widehat{B}^TA_1$ that are linearly
dependent. Denote $C_p$ the submatrix consisting of these $p$
columns. Then the Gram matrix of $C_p$, $G_{pp}=C_p^TC_p \in
R^{p\times p}$, is singular. Since all diagonal entries of $G_{pp}$
are 1's, and the absolute value of  off-diagonal entries are less
than or equal to $\mu(\widehat{B}^TA_1)$. Under either condition (i)
or (ii) of the theorem, we have
$$\alpha^*_{A_2}(A_1)\leq  \frac{1}{\mu^{**}_{A_2}(A_1)} \leq
\frac{1}{\mu(B^TA_1)} \textrm{ for any } B\in F. $$  In particular,
we have
\begin{equation} \label{pp11}  \alpha^*_{A_2}(A_1) \leq  \frac{1}{\mu(\widehat{B}^TA_1)}\leq
\textrm{Spark}(\widehat{B}^TA_1) -1 =p-1.
\end{equation}
Since $G_{pp}$ is a $p\times p$ matrix, in each row of $G_{pp}$,
there are at most $\alpha^*_{A_2}(A_1) = \alpha(\widehat{B}^T A_1)$
entries whose absolute values are equal to $\mu(\widehat{B}^TA_1)$,
and the absolute values of
  the remaining $(p-1-\alpha^*_{A_2}(A_1))$
entries are less than or equal to $\mu^{(2)}(\widehat{B}^TA_1)$. By
the singularity of $G_{pp}$, we know that $\lambda=0$ is an
eigenvalue of $G_{pp}$. By Lemma \ref{Thm1.5}, there exist two rows
of $G_{pp}$, say, the $i$th row and the $j$th row ($i\neq j$),
satisfying that
\begin{eqnarray}\label{eqnBrau}
|0-G_{ii}|\cdot|0-G_{jj}|\leq\Delta_i\cdot\Delta_j=\sum_{t=1,t\neq i}^p|c_i^Tc_t|\cdot\sum_{t=1,t\neq j}^p|c_j^Tc_t|.
\end{eqnarray}
By the definitions of coherence rank and subcoherence rank, if there
are $\alpha^*_{A_2}(A_1) (=\alpha(\widehat{B}^T A_1)) $ entries
whose absolute values are $\mu(\widehat{B}^TA_1)$ in the $i$th row,
 then for the $j$th row, there are at most $\beta(\widehat{B}^TA_1)$ entries whose absolute
 values are $\mu(\widehat{B}^TA_1)$. And the absolute values of the remaining
  entries in either row are less than or equal to $\mu^{(2)}(\widehat{B}^TA_1)$.
Therefore, from (\ref{eqnBrau}), we have that
\begin{eqnarray}\label{eqnrelax1}
1&\leq&[\alpha^*_{A_2}(A_1)\mu(\widehat{B}^TA_1)+(p-1-\alpha^*_{A_2}(A_1))\mu^{(2)}(\widehat{B}^TA_1)]\cdot\\
&
&[\beta(\widehat{B}^TA_1)\mu(\widehat{B}^TA_1)+(p-1-\beta(\widehat{B}^TA_1))\mu^{(2)}(\widehat{B}^TA_1)]\nonumber.
\end{eqnarray}
Let $p^*=\textrm{Spark}_{A_2}^*(A_1)$.  Since
$\textrm{Spark}_{A_2}^*(A_1)$ is the supremum of $\textrm{Spark}(B^T
A_1)$ over   $ F, $  we have $p\leq p^*.$
  Thus it follows from
(\ref{eqnrelax1}) that
\begin{eqnarray}\label{eqnrelax2}
1&\leq&[\alpha^*_{A_2}(A_1)\mu(\widehat{B}^TA_1)+(p^*-1-\alpha^*_{A_2}(A_1))\mu^{(2)}(\widehat{B}^TA_1)]\cdot\\
&
&[\beta(\widehat{B}^TA_1)\mu(\widehat{B}^TA_1)+(p^*-1-\beta(\widehat{B}^TA_1))\mu^{(2)}(\widehat{B}^TA_1)]\nonumber.
\end{eqnarray}
By the definition of $\beta^*_{A_2}(A_1)$, we have
$\beta(\widehat{B}^TA_1)\leq\beta_{A_2}^*(A_1). $ This, together
with $\mu (\widehat{B}^T A) \geq \mu^{(2)} (\widehat{B}^T A_1),$
implies that
\begin{eqnarray*}
  & & \beta( \widehat{B}^TA_1)\mu(\widehat{B}^TA_1)+(p^*-1-\beta(\widehat{B}^TA_1))\mu^{(2)}(\widehat{B}^TA_1) \\
   & \leq &
   \beta_{A_2}^*(A_1)\mu(\widehat{B}^TA_1)+(p^*-1-\beta_{A_2}^*(A_1))\mu^{(2)}(\widehat{B}^TA_1).
\end{eqnarray*}
Combining  (\ref{eqnrelax2}) with the inequality above yields
\begin{eqnarray}\label{eqnrelax3}
1&\leq&[\alpha^*_{A_2}(A_1)\mu(\widehat{B}^TA_1)+(p^*-1-\alpha^*_{A_2}(A_1))\mu^{(2)}(\widehat{B}^TA_1)]\cdot  \nonumber \\
&
&[\beta_{A_2}^*(A_1)\mu(\widehat{B}^TA_1)+(p^*-1-\beta_{A_2}^*(A_1))\mu^{(2)}(\widehat{B}^TA_1)].
\end{eqnarray}
Note that $$ \beta^*_{A_2}(A_1) \leq \alpha^*_{A_2}(A_1)\leq p-1\leq
p^*-1, ~ \mu(\widehat{B}^TA_1) \leq  \mu^{**}_{A_2}(A_1), ~
\mu^{(2)}(\widehat{B}^TA_1) \leq \mu^{**(2)}_{A_2}(A_1) .$$ So from
(\ref{eqnrelax3}), we obtain
\begin{eqnarray*}\label{eqnrelax4}
1 & \leq &[\alpha^*_{A_2}(A_1)\mu^{**}_{A_2}(A_1)+(p^*-1-\alpha^*_{A_2}(A_1))\mu^{**(2)}_{A_2}(A_1)]\cdot\nonumber\\
& &
[\beta_{A_2}^*(A_1)\mu^{**}_{A_2}(A_1)+(p^*-1-\beta_{A_2}^*(A_1))\mu^{**(2)}_{A_2}(A_1)].
\end{eqnarray*}
Denote by $\bar{\mu}^*:=\mu^{**}_{A_2}(A_1)-\mu^{**(2)}_{A_2}(A_1)$.
The above inequality can be written as
\begin{small}
\begin{eqnarray}\label{eqnrelax5}
\left[(p^*-1)\mu^{**(2)}_{A_2} (A_1)\right]^2+(p^*-1)(\alpha^*_{A_2}
(A_1)+\beta_{A_2}^* (A_1)) \bar{\mu}^* \mu^{**(2)}_{A_2}
(A_1)+\alpha^*_{A_2} (A_1)\beta_{A_2}^* (A_1)(\bar{\mu}^*)^2 \geq 1.
\end{eqnarray}
\end{small}
By the definition of $\mu^{**(2)}_{A_2}(A_1),$ we know that
$\mu^{**(2)}_{A_2}(A_1)\geq 0$. We now prove that
$\mu^{**(2)}_{A_2}(A_1)>0. $ In fact, if $\mu^{**(2)}_{A_2}(A_1)=0,$
then the quadratic inequality (\ref{eqnrelax5}) becomes
$$\alpha^*_{A_2}(A_1)\beta_{A_2}^*(A_1)\left(\mu^{**}_{A_2}(A_1)\right)^2\geq 1,$$ which contradicts to either
condition (i) or condition (ii) of the theorem. Thus
$\mu^{**(2)}_{A_2}(A_1)$ must be positive.  Consider the following
quadratic equation in variable $t: $
\begin{eqnarray*}
h(t):=t^2+t(\alpha^*_{A_2}(A_1)+\beta_{A_2}^*(A_1))\bar{\mu}^*+\alpha^*_{A_2}(A_1)\beta_{A_2}^*(A_1)(\bar{\mu}^*)^2-1=0
\end{eqnarray*}
which has only one positive root under conditions (i) and (ii). This
positive root is given by
\begin{eqnarray*}
t^*=\frac{-(\alpha^*_{A_2}(A_1)+\beta_{A_2}^*(A_1))\bar{\mu}^*+\sqrt{\rho}}{2},
\end{eqnarray*}
where
$\rho=(\alpha^*_{A_2}(A_1)-\beta_{A_2}^*(A_1))^2(\bar{\mu}^*)^2+4$.
Let $\gamma =(p^*-1)\mu^{**(2)}_{A_2}(A_1). $ The inequality
(\ref{eqnrelax5}) shows that $h(\gamma)\geq 0.$ Thus $\gamma \geq
t^*,$ that is,
$$
(p^*-1)\mu^{**(2)}_{A_2}(A_1) \geq
\frac{-(\alpha^*_{A_2}(A_1)+\beta_{A_2}^*(A_1))\bar{\mu}^*+\sqrt{\rho}}{2}.
$$ Therefore,
$$
 \textrm{Spark}^*_{A_2}(A_1) = p^* \geq 1+\frac{\sqrt{\rho}-(\alpha^*_{A_2}(A_1)+\beta_{A_2}^*(A_1))
 \bar{\mu}^*}{2\mu^{**(2)}_{A_2}(A_1)},  $$
 as desired.\\

By Theorem  \ref{UniONSSpark} and Theorem \ref{Thmsparkstar}, we
immediately have the next uniqueness condition.

\begin{Cor} Consider the system (\ref{system}) where $A_1\in R^{m\times n_1},$  $A_2\in R^{m\times n_2},$
 and  $ m<n_1. $ Under the
same condition of Theorem \ref{Thmsparkstar}.  If there exists a
solution $(x,y)$ to the system (\ref{system}) satisfying that
$$
\|x\|_0<\frac{1}{2} \varphi^* =:
\frac{1}{2}\left(1+\frac{\sqrt{\rho}-(\alpha^*_{A_2}(A_1)+\beta_{A_2}^*(A_1))
 \bar{\mu}^*}{2\mu^{**(2)}_{A_2}(A_1)}\right),
$$
 then $x$ is the
unique sparsest x-part solution to the system (\ref{system}).

\end{Cor}

 The above corollary may also provide a tighter lower bound of $\textrm{Spark}_{A_2}^*(A_1)$ than  Theorems
 \ref{Thm1.4}
under some conditions, as indicated by the following proposition.

\begin{Propn} \label{Prp1.1}
Let $\varphi^*$ be a lower bound of $Spark_{A_2}^*(A_1)$ given in
Theorem \ref{Thmsparkstar}. Assume that $\alpha^*_{A_2}(A_1)=1$ and
$\alpha^*_{A_2}(A_1)<\frac{1}{\mu_{A_2}^{**}(A_1)}$. If
$\mu_{A_2}^{**(2)}(A_1) <  \mu^*_{A_2}(A_1)(1-\bar{\mu}^*)
   $ where $ \bar{\mu}^*=
\mu_{A_2}^{**}(A_1)-\mu_{A_2}^{**(2)}(A_1), $  we have
$
\varphi^*
 >1+\frac{1}{\mu^*_{A_2}(A_1)}.
$
\end{Propn}

{\it Proof.} Under condition
$\alpha^*_{A_2}(A_1)<\frac{1}{\mu_{A_2}^{**}(A_1)}$, by Theorem
(\ref{Thmsparkstar}) we get the following lower bond of
$\textrm{Spark}^*_{A_2}(A_1): $
\begin{eqnarray}\label{eqn1.30}
  \varphi^*= 1+ \frac{\sqrt{\rho}-(\alpha^*_{A_2}
(A_1)+\beta_{A_2}^*(A_1))\bar{\mu}^*}{2\mu^{**(2)}_{A_2}(A_1)}.
\end{eqnarray}
By (\ref{alpha-beta}),  we see that $\alpha^*_{A_2}(A_1)=1$ implies
that $\beta^*_{A_2}(A_1)=1$. Thus (\ref{eqn1.30}) is reduced to $
\varphi^* -1=    \frac{1-\bar{\mu}^*}{\mu^{**(2)}_{A_2}(A_1)}.$ Note
that
\begin{eqnarray*}
\frac{1-\bar{\mu}^*}{\mu^{**(2)}_{A_2}(A_1)}&=&\frac{1}{\mu^*_{A_2}(A_1)}+
\left(\frac{1-\bar{\mu}^{*}}
{\mu^{**(2)}_{A_2}(A_1)}-\frac{1}{\mu^*_{A_2}(A_1)}\right) \\
&=&\frac{1}{\mu^*_{A_2}(A_1)}+
\frac{\mu^*_{A_2}(A_1)(1-\bar{\mu}^*)- \mu_{A_2}^{**(2)}(A_1)     }{\mu^{**(2)}_{A_2}(A_1)\mu^*_{A_2}(A_1)},\\
\end{eqnarray*}
Thus if $\mu_{A_2}^{**(2)}(A_1) <  \mu^*_{A_2}(A_1)(1-\bar{\mu}^*),$
  we must have
$ \varphi^*>1+ \frac{1}{\mu^*_{A_2}(A_1)}. $ \\

The discussion in this section demonstrates that the concepts
introduced in this section such as maximal scaled coherence rank and
subcoherence rank, and minimal/maximal scaled mutual coherence are
quite  useful in the development of uniqueness criteria for the
  $\ell_0$-minimization problem (\ref{PPPP}).



\section{Conclusion}
In this paper, we have established several uniqueness conditions for
the solution to a class of  $\ell_0$-minimization problems which
 seek sparsity only for part of the variables of the problem. This
problem includes several important sparsity-seeking models as
special cases. To obtain uniqueness conditions, several concepts
such as maximal/minimal scaled coherence and maximal scaled
coherence rank, maximal scaled spark have been introduced in this
paper. Also the $l_p$-induced norm has been defined and used to
establish a sufficient condition for the uniqueness of the solution
to the underlying $\ell_0$-minimization problems as well.


\begin{thebibliography}{999}

\bibitem{BSV2013} A. S. Bandeira and K. Scheinberg and L. N. Vicente,
On partial sparse recovery, Submitted to \emph{IEEE Signal
Processing letters}, 2013
\bibitem{BSV2012} A. S. Bandeira and K. Scheinberg and L. N. Vicente,
Computation of sparse low degree interpolating polynomials and their
application to derivative-free optimization, \emph{Mathematical
Programming}, 134 (2012), 223-257.
\bibitem{AB1946} A. Brauer, Limits for the characteristic roots of a matrix, \emph{Duke Math. J.},
13 (1946), 387-395.
\bibitem{BED2009} A. Bruckstein and M.Elad and D. Donoho, From sparse solutions of systems of equations
to sparse modeling of signals and images, \emph{SIAM Review}, 51(2009), 34-81.
\bibitem{DDEK2012} M. Davenport and M. Duarte and Y. Eldar and G. Kutyniok, \emph{Introduction to
Compressive Sensing}, in Compressive Sensing: Theory and
Applications (Eldar and Kutyniok eds.), University of Cambridge,
2012.
\bibitem{DE2003} D. Donoho and M. Elad, Optimally sparse representation in general (nonorthogonal)
dictionaries via $L^1$ minimization, \emph{Proc. Nat. Acad. Sci.},
100 (2003), 2197-2202.
\bibitem{DH2001} D. Donoho and X. Huo, Uncertainty principles and ideal atomic decompositions,
\emph{IEEE Transactions on Information Theory}, 47 (2001),
2845-2862.
\bibitem{DOLI2011} A. Doostan and H. Owhadi and A. Lashgari and G. Iaccarino, Non-adapted sparse
 approximation of PDEs with stochastic inputs, \emph{Journal of Computational Physics}, 230 (2011), 3015-3034.
\bibitem{ME2010} M. Elad, \emph{Sparse and Redundant Representations: From Theory to Applications
in Signal and Image Processing}, Springer, New York, 2010.
\bibitem{EK2012} Y. Eldar and G. Kutyniok, \emph{Compressive Sensing: Theory and Applications},
Cambridge University Press, 2012.
\bibitem{FNW2007} M. Figueiredo and R. Nowak and  S. Wright, Gradient projection for sparse reconstruction:
application to compressed sensing and other inverse problems,
\emph{IEEE Journal of Selected Topics in Signal Processing}, 1
(2007), 586-597.
\bibitem{FPRU2010} S. Foucart and A. Pajor and H. Rauhutc and T. Ullrich, The Gelfand widths of
$l_p$-balls for $0<p<1$, \emph{Journal of Complexity}, 26 (2010),
629-640.
\bibitem{J2010} L. Jacques, A short note on compressed sensing with partially known signal support,
\emph{Signal Processing}, 90 (2010), 3308-3312.
\bibitem{LSLD2005} M. Lustig and D. Donoho and J. Pauly, Sparse MRI: the application of compressed
sensing for rapid MR imaging, \emph{Magnetic Resonance in Medicine}
58 (2007), 1182–1195.
\bibitem{SM1998} S. Mallat, \emph{A Wavelet Tour of Signal Processing}, Academic Press, 1998.
\bibitem{N1995} B. Natarajan, Sparse approximate solutions to linear systems, \emph{SIAM Journal on Computing},
24 (1995), 227-234.
\bibitem{TM2001} D. Taubman and M. Marcellin, \emph{JPEG2000: Image Compression Fundamentals, Standards and Practice},
Kluwer Academic, 2001.
\bibitem{VL2010} N. Vaswani and W. Lu, Modified-CS: Modifying compressive sensing for problems with partially known
support, \emph{IEEE Transactions on Signal Processing}, 58 (2010),
4595-4607.
\bibitem{YZ012} Y.B. Zhao, New and improved conditions for uniqueness of sparsest solutions of underdetermined linear
systems, \emph{Applied Mathematics and Computation}, 224 (2013),
58-73.
\bibitem{YZ2013} Y.B. Zhao, RSP-Based analysis for sparest and least $l_1$-norm solutions to underdetermined linear
systems, \emph{IEEE Transactions on Signal Processing}, 61 (2013),
no. 22, 5777-5788.
\bibitem{ZL2012} Y.B. Zhao and D. Li, Reweighted $l_1$-minimization for sparse solutions to underdetermined linear
system, \emph{SIAM Journal on Optimization}, 22 (2012), no. 3,
1065-1088.


\end{thebibliography}
\end{document}